\documentclass[%
 aip,
 rsi,
 amsmath,amssymb,
 reprint,%
longbibliography
]{revtex4-1}

\usepackage{graphicx}
\usepackage{dcolumn}
\usepackage{bm}

\usepackage[utf8]{inputenc}
\usepackage[T1]{fontenc}
\usepackage{mathptmx}
\usepackage{etoolbox}
\usepackage[dvipsnames]{xcolor}

\newcommand{\cecoin}{CeCoIn$_5$}
\newcommand{\sro}{Sr$_2$RuO$_4$}
\newcommand{\upt}{UPt$_3$}
\newcommand{\I}{\mathrm{i}}

\makeatletter
\def\@email#1#2{%
 \endgroup
 \patchcmd{\titleblock@produce}
  {\frontmatter@RRAPformat}
  {\frontmatter@RRAPformat{\produce@RRAP{*#1\href{mailto:#2}{#2}}}\frontmatter@RRAPformat}
  {}{}
}%
\makeatother
\begin{document}

\preprint{AIP/123-QED}

\title{Microwave resonator for measuring time-reversal symmetry breaking at cryogenic temperatures}
\author{T. Chouinard}
\affiliation{ 
Department of Physics, Simon Fraser University, Burnaby, British Columbia V5A 1S6, Canada}
\author{D. M. Broun}%
\affiliation{ 
Department of Physics, Simon Fraser University, Burnaby, British Columbia V5A 1S6, Canada
}%

\date{\today}

\begin{abstract}
We present a microwave-frequency method for measuring polar Kerr effect and spontaneous time-reversal symmetry breaking (TRSB) in unconventional superconductors.  While this experiment is motivated by work performed in the near infrared using zero-loop-area Sagnac interferometers, the microwave implementation is quite different, and is based on the doubly degenerate modes of a TE$_{111}$ cavity resonator, which act as polarization states analogous to those of light.  The resonator system has \emph{in-situ} actuators that allow quadrupolar distortions of the resonator shape to be controllably tuned, as these compete with the much smaller perturbations that arise from TRSB.  The most reliable way to the detect the TRSB signal is by interrogating the two-mode resonator system with circularly polarized microwaves, in which case the presence of TRSB shows up unambiguously as a difference between the forward and reverse transmission response of the resonator --- i.e., as a breaking of reciprocity.  We illustrate and characterize a coupler system that generates and detects circularly polarized microwaves, and then show how these are integrated with the TE$_{111}$  resonator, resulting in a dilution refrigerator implementation with a base temperature of 20~mK.  We show test data on yttrium-iron-garnet (YIG) ferrite and the van der Waals ferromagnet CrGeTe$_3$ as an illustration of how the system operates, then present data showing system performance under realistic conditions at millikelvin temperatures.
\end{abstract}

\maketitle

\section{\label{sec:Introduction}Introduction}

Time-reversal symmetry breaking (TRSB) is a subtle and elusive broken-symmetry state found in some unconventional superconductors, such as \sro\ \cite{Luke:1998bo,Kidwingira:2006p807,Xia:2006p2} and \upt.\cite{Luke:1993fc,Sauls:1994p138,Joynt2002,Strand:2009p137,Strand:2010p42,Schemm:2014fv} In the case of \sro\ \cite{Mackenzie:2017do} TRSB is believed to be associated with the existence of a pairing state such as chiral \mbox{$p$-wave} ($p_x \pm \I p_y$); chiral $d$-wave ($d_{xz} \pm \I d_{yz}$); or exotic nonchiral states such as  $d_{x^2 -y^2} \pm \I g_{xy(x^2 - y^2)}$,\cite{Kivelson:2020} $d_{x^2 - y^2} \pm \I s$,\cite{Romer:2020} or $d_{xy} \pm \I s$.\cite{Romer:2021} In the case of \upt\ many experiments are well described by the $E_{2u}$ odd-parity triplet state, whose orbital part takes the chiral $f$-wave form,  $(k_x^2 - k_y^2)k_z \pm 2 \I k_x k_y k_z $.\cite{Sauls:1994p138,Joynt2002} However, the interpretation of experiments reporting TRSB is controversial and remains an active topic of investigation, with the possibility that in some cases TRSB may be arising from the interplay between superconductivity and disorder.\cite{Andersen:2024}  Insights obtained from the development of new techniques for probing TRSB are therefore highly desirable.

\begin{figure}[b]
\includegraphics[width = 0.84 \columnwidth]{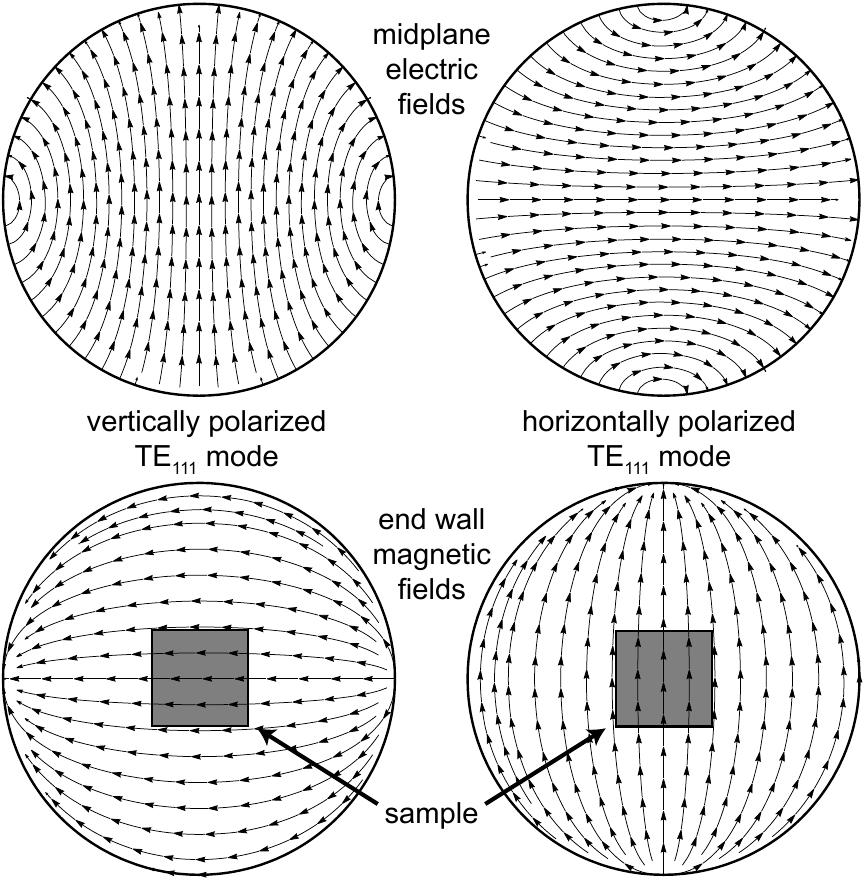}
\caption{The vertical (left column) and horizontal (right column) polarizations of the doubly degenerate TE$_{111}$ mode of a cylindrical cavity.  The electric field is purely transverse, and has maximum magnitude in the mid plane of the resonator, as shown in the top row of plots.  The magnetic field forms loops that extend in the $z$ direction, but is purely transverse at the end walls, as shown in the bottom row.  The sample, indicated by the square, is located in the center of an end wall, at a maximum of the transverse magnetic field. When the modes are superimposed to form circular polarizations, the field at the sample rotates continuously, clockwise or counterclockwise.}
\label{fig:TE111_modes}
\end{figure}

The polar Kerr effect is an experimental observable that definitively indicates time-reversal symmetry has been broken. The Kerr angle, $\theta_K$, is proportional to the nonreciprocal phase shift, $\phi_\mathrm{nr}$, between left- and right-circularly polarized EM waves reflected from the surface of a sample with a component of magnetization perpendicular to the surface.  This occurs when the electromagnetic tensors (e.g., $\sigma$, $\epsilon$, $Z_s$) contain a skew-symmetric off-diagonal component. 

\begin{figure*}[thb]
\includegraphics[width = 0.95 \textwidth]{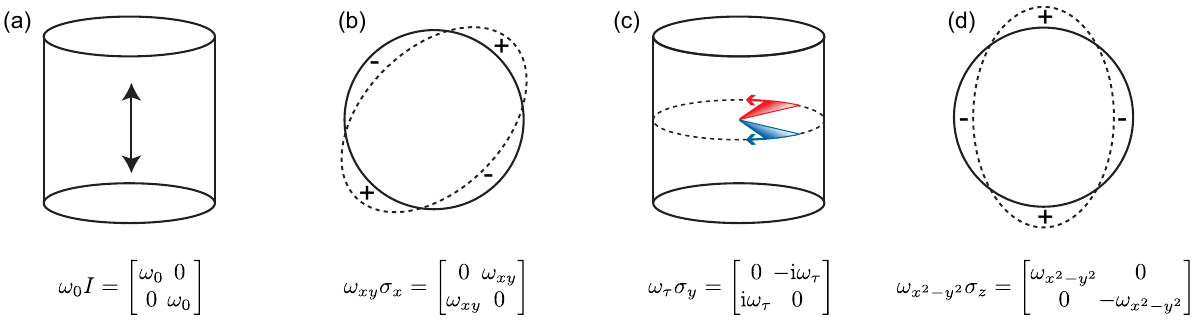}
\caption{Schematic representation of the four independent types of perturbation to the degenerate TE$_{111}$ modes of a cylindrical cavity resonator, along with the corresponding matrix perturbation in the $z$ basis. (a) common-mode perturbations, which affect both TE$_{111}$ polarizations equally; (b) $xy$-type quadrupolar distortions; (c) time-reversal-symmetry-breaking perturbations, which have opposite effect on left- and right-circular polarizations; and (d) ($x^2 - y^2$)-type quadrupolar distortions.}
\label{fig:perturbations}
\end{figure*}

Observations of Kerr effect and spontaneous TRSB in superconcductors have been made by Kapitulnik and coworkers using a modified zero-loop Sagnac interferometer operating in the near infrared.\cite{Xia:2006p36,Xia:2006p2,Kapitulnik:2009p41,Kapitulnik:2009p34,Schemm:2014fv}  The magnitude of Kerr angle detected in the broken-time-reversal-symmetry superconducting states is tiny, of the order of tens of nanoradians. This is testament to how well their modified Sagnac interferometer rejects spurious common-mode signals, and is an indication of the difficulty faced by other experimental methods.

Inspired by the Sagnac experiments, we present a new means for measuring $\theta_K$, operating in the microwave range.  Part of the motivation for carrying out these experiments at microwave frequencies is that unconventional superconductors for which spontaneous TRSB and Kerr effect have been observed, such as \sro\ and \upt\, have superconducting $T_c$ of the order of 1~K~$\approx 20$~GHz.  We therefore expect spectroscopic signatures (i.e., changes in the optical conductivity spectrum, $\sigma(\omega)$, due to the onset of superconductivity) to be strong in the microwave range.

\section{\label{sec:Kerr_effect}Kerr effect at microwave frequencies}

One of the most important reasons to study TRSB using measurements of polar Kerr effect is that the Kerr angle, \mbox{$\theta_K = \phi_\mathrm{nr}/2$}, is inherently sensitive to nonreciprocal effects --- it is only nonzero when there is a difference in phase between left- and right-circulalry polarized waves (LCP and RCP) reflected from a surface.  This allows it to distinguish between true Faraday effects due to circular birefringence, and the Faraday rotation that occurs when linearly polarized waves propagate through a linearly birefringent material.

 The starting point for analyzing the electromagnetic Kerr response is the Argyres formula for Kerr angle,\cite{Argyres55}
\begin{equation}
    \theta_K = - \mathrm{Im}\left\{\frac{\tilde n_+ - \tilde n_-}{\tilde n_+ \tilde n_- -1} \right\}\;,
    \label{eq:Argyres}
\end{equation}
where $\tilde n_\pm$ are the complex indices of refraction for left- and right-circularly polarized waves.  A nonzero Kerr angle implies that $\tilde n_+ \ne \tilde n_-$, indicating that time-reversal symmetry has been broken.  In the microwave frequency range it is more common to represent the electrodynamic response of the material by its complex surface impedance, \mbox{$Z_s = R_s + \mathrm{i} X_s$}. Surface impedance can be related to refractive index via \mbox{$\tilde n = Z_0/Z_s$}, where \mbox{$Z_0 = \sqrt{\mu_0/\epsilon_0} = 377~\Omega$} is the impedance of free space.  As shown in Ref.~\onlinecite{Chouinard.2025A}, when this is substituted into the Argyres formula, we obtain a microwave-frequency expression for the Kerr angle:
\begin{equation}
    \theta_K = \mathrm{Im}\left\{\frac{Z_s^+ - Z_s^-}{Z_0 - Z_s^+ Z_s^-/Z_0} \right\} \approx \mathrm{Im}\left\{\frac{Z_s^+ \!-\! Z_s^-}{Z_0} \right\} = \frac{X_s^+ \!-\! X_s^-}{Z_0}.
\end{equation}
That is, the Kerr angle is proportional to the difference in surface reactance for left- and right-circularly polarized microwaves. Surface reactance measurements are typically carried out using cavity perturbation measurements,\cite{Ormeno:1997p342,Huttema:2006p344} meaning that Kerr angle should be visible in the frequency shift of a microwave cavity. The actual method for detecting $\theta_K$ is more complicated, and we describe it below.

\section{\label{sec:Experimental_Method}Experimental method}

A direct transposition of the optical Sagnac interferometer to the microwave range is not realistic, principally because the much longer microwave wavelength makes it impossible to avoid coherent interference from spurious reflections. Instead, we have recently proposed that a different type of interferometer --- the well-known microwave cavity resonator --- can instead be used to measure polar Kerr effect.\cite{Chouinard.2025A}   The microwave resonator must support a pair of degenerate modes, such as the TE$_{111}$ modes of a cylindrical cavity.  As shown in Fig.~\ref{fig:TE111_modes}, the pair of modes acts analogously to the horizontal and vertical polarization states familiar from optics.  We refer to these states as the $z$ basis and label them $|0\rangle =  \left(\begin{smallmatrix}1\\0\end{smallmatrix}\right)$ for the verical polarization and $|1\rangle = \left(\begin{smallmatrix}0\\1\end{smallmatrix} \right)$ for the horizontal polarization.  A sample of the material of interest is positioned inside the resonator to interact with the EM fields of these modes: in our case, a small single-crystal sample is placed near one of the end walls of the resonator, where the transverse component of the microwave $H$ field is maximum.  If the sample is a material that strongly breaks time-reversal symmetry (e.g., a soft, magnetized ferromagnet) without introducing shape distortions, the eigenmodes of the resonator will be circularly polarized superpositions of the linear polarization states shown in Fig.~\ref{fig:TE111_modes}: $|\pm\rangle = \tfrac{1}{\sqrt{2}}\big(|0\rangle \pm \I |1\rangle \big)$.  These circularly polarized modes correspond physically to a transverse microwave $H$ field that rotates clockwise or counterclockwise at the center of the end wall of the resonator, where the sample is located.  In the limit of strong TRSB, a measurement of the frequency splitting between the $|+\rangle$ and $|-\rangle$ modes would, on its own, provide a direct measurement of the Kerr angle.

For measurements on superconductors such as \sro\ and \upt, the degree of TRSB is very weak, and a more mathematically involved treatment is required.  As shown in Ref.~\onlinecite{Chouinard.2025A}, the pair of degenerate TE$_{111}$ modes, resonant at frequency $\omega_0$, forms a $2 \times 2$ subspace, perturbations to which are spanned by the set of Pauli matrices, $\sigma_i$. That is, 
\begin{equation}
    H = H_0 + H_1 = \omega_0 I + \omega_{xy} \sigma_x + \omega_\tau \sigma_y + \omega_{x^2 - y^2} \sigma_z\;,
\end{equation}
where $H$ is the system matrix.  As illustrated in Fig.~\ref{fig:perturbations}, the identity matrix corresponds to common-mode perturbations, such as uniform resonator stretch. Two of the other perturbations correspond to shape distortions of the resonator:
\begin{align}
    \omega_{xy} \sigma_x & = \left[\begin{array}{cc}0 & \omega_{xy} \\\omega_{xy} & 0\end{array}\right]\;,\\
    \omega_{x^2-y^2} \sigma_z & = \left[\begin{array}{cc}\omega_{x^2-y^2} & 0 \\0 & -\omega_{x^2-y^2}\end{array}\right]\;.
\end{align}
It is the $\sigma_y$ term that breaks time-reversal symmetry, and is quantified by $\omega_\tau$: 
\begin{equation}
    \omega_\tau \sigma_y = \left[\begin{array}{cc}0 & - \I \omega_\tau \\ \I \omega_\tau & 0\end{array}\right]\;.
\end{equation}

When all three perturbations are present, the problem is analogous to a spin-$\tfrac{1}{2}$ particle in a magnetic field, with \mbox{$H_1 = \vec \Omega \cdot \vec \sigma$}, where $\vec \Omega = (\omega_{xy},\omega_\tau,\omega_{x^2 - y^2})$ and $\vec \sigma = (\sigma_x,\sigma_y,\sigma_z)$.  The frequency splitting therefore takes the form of a Dirac spectrum:
\begin{align}
    \begin{split}
    \omega &= \omega_0 \pm | \Omega | \\
        & = \omega_0 \pm \sqrt{\smash{\omega^2_{xy} + \omega_\tau^2 + \omega_{x^2 - y^2}^2} \vphantom{|}}\;.
    \end{split}
\end{align}
The individual perturbations add in quadrature, which in practice means that weak TRSB will be overwhelmed by shape perturbations, which inevitably are larger.  Nevertheless, it is useful to be able to control the shape perturbations, so the resonator system described here uses a set of eight actuators to control the quadrupolar shape distortions \emph{in situ}.

Instead of directly detecting TRSB from the frequency splitting, we have shown in Ref.~\onlinecite{Chouinard.2025A} that a better way to obtain this information is to interrogate the resonator using circularly polarized microwaves.  The end result of that analysis is that in the presence of quadrupolar distortion, the transmission amplitudes through the pair of perturbed TE$_{111}$ resonator modes is given by \mbox{$A_\pm  = \tfrac{1}{2} \big(1 \pm \omega_\tau/|\Omega| \big)$} in the forward direction, and 
\mbox{$\overline A_\pm  = \tfrac{1}{2} \big(1 \mp \omega_\tau/|\Omega| \big)$} in the reverse section.  This immediately reveals the breaking of reciprocity, and shows that the TRSB parameter $\omega_\tau$ can be accessed via combinations such as \begin{equation}
    \frac{A_+}{A_-}\times \frac{\overline A_-}{\overline A_+} = \frac{\left(1 + \omega_\tau/|\Omega|\right)^2}{\left(1 - \omega_\tau/|\Omega|\right)^2} \approx 1 + 4 \omega_\tau/|\Omega|\;,
    \label{eq:ratiometric_measure}
\end{equation} 
where the final result applies when $\omega_\tau \ll |\Omega|$.  In the next section we describe the implementation of this system, and then go on to show it in operation.

\section{Experimental apparatus}

\subsection{Circularly polarized resonators and interferometers}

Cavity resonators supporting circularly polarized modes date back 70~years to Dresselhaus, Kip and Kittel, \cite{Dresselhaus:1955wm} who used a clever waveguide implementation operating in a reflection geometry to separately probe the left- and right-circularly polarized response of semiconductors in cyclotron-resonance experiments in strong magnetic field, thereby distinguishing between electrons and holes.  In the Dresselhaus experiment, quadrupolar-type distortions were tuned using a rotatable dielectric vane inserted from the far end of their resonator, which was possible because they performed a single-ended reflection measurement from the opposite end of the resonator.  In a more recent implementation of the Dresselhaus experiment the input waveguide/circular-polarizer assembly is replaced by a hybrid coupler and coaxial cables,\cite{Arakawa.2019,Roppongi.2024} again allowing the resonator to be probed in reflection by LCP and RCP microwaves.  This has been used to study materials ranging from YIG ferrites to quantum Hall-effect devices,  but not to measure Kerr angle.\cite{Arakawa.2022}  In other experiments, the polar Kerr effect has been measured at sub-THz frequencies using a modified Martin–Puplett interferometer, albeit with milliradian resolution, which is not sufficiently sensitive to detect spontaneous TRSB in unconventional superconductors. \cite{Moshe.2024}

\begin{figure}[t]
\includegraphics[width = 0.8 \columnwidth]{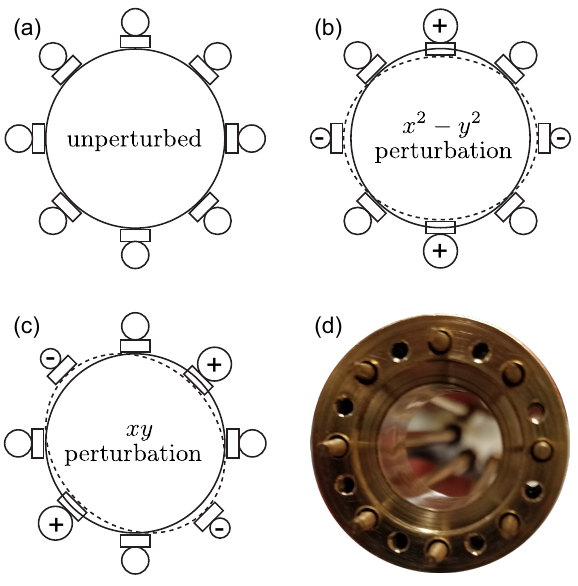}
\caption{Perturbations to (a) an undistorted cylindrical resonator are controllably introduced via a set of eight drive rods and contact pads to generate (b) an $x^2 - y^2$ distortion and (c) an $xy$ distortion of the resonator walls.  (d) A photograph of the BeCu resonator walls showing a controlled and reversible pattern of distortion.}
\label{fig:tunable_quadrupoles}
\end{figure}

\subsection{Deformable cavity resonator}

\begin{figure*}[t]
\includegraphics[width = 0.95 \textwidth]{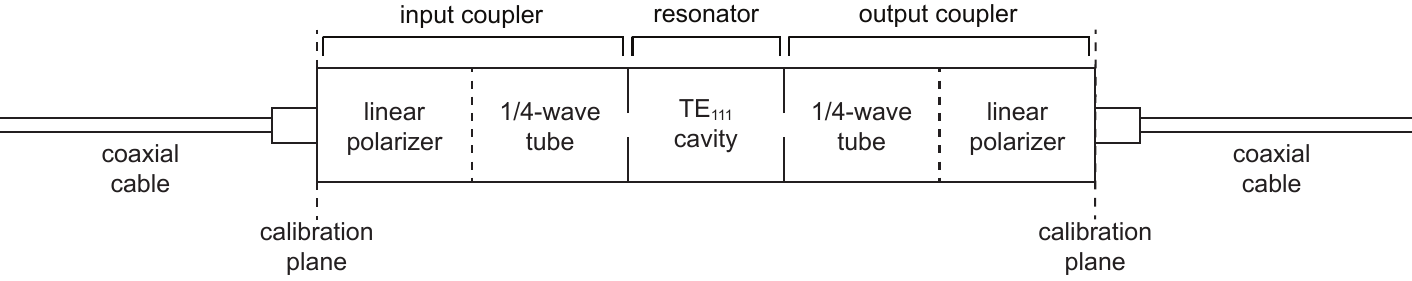}
\caption{Schematic of the TRSB resonator system.  From left to right, microwaves are: launched from an input coaxial cable; linearly polarized; converted to circularly polarized microwaves by a $\tfrac{1}{4}$-wave tube; coupled into and out of the TE$_{111}$ modes of a cavity resonator; passed through a second $\tfrac{1}{4}$-wave tube; and then a second linear polarizer, with that linear polarization coupled into the output coaxial cable.  The combined effect of the pair of linear polarizers and $\tfrac{1}{4}$-wave tubes is to interrogate the resonator with right-circularly polarized microwaves in the forward direction, and left-circularly polarized microwaves in the reverse direction.  Any perturbation to the resonator that breaks time-reversal symmetry is revealed as a breaking of reciprocity.}
\label{fig:resonator_schematic}
\end{figure*}

In our implementation, the experiment centers around a cylindrical cavity resonator with deformable side walls, as shown in Fig.~\ref{fig:tunable_quadrupoles}.  The resonator supports a pair of nominally degenerate TE$_{111}$ modes resonant at 16--17~GHz, resulting from a resonator of nominal height 17-18~mm and diameter 12-12.7~mm. The side walls are made from thin-walled BeCu tubing of wall thickness 100~$\mu$m. As-purchased, the BeCu tubing is in solution-annealed form.  After cutting approximately to size, the BeCu tubing undergoes a standard age-hardening process for 2~hours at 300~$^\circ$C in 1~atm Ar gas to improve its mechanical properties, in particular its plastic yield stress.  After annealing, the BeCu tubing is highly elastic, and no plastic deformation of the resonator takes place during normal operation.

To enable \emph{in-situ} tuning of the resonator shape, the resonator side walls are surrounded by a set of eight PTFE contact pads, as shown in Fig.~\ref{fig:tunable_quadrupoles}a.  The $x^2 - y^2$ (Fig.~\ref{fig:tunable_quadrupoles}b) and $xy$ (Fig.~\ref{fig:tunable_quadrupoles}b) quadrupoles can be separately controlled using a set of eight tapered tuning screws, which independently push on each of the PTFE pads through a brass coupling.  A given quadrupolar distortion is activated by partially engaging one diametrically opposite pair of tuning screws while simultaneously partially disengaging the diametric pair at 90$^\circ$ to it.  In this way, the quadrupolar distortions can easily be tuned over a range of several hundred MHz, more than sufficient to tune out any deviation from cylindrical symmetry introduced by the sample and sample holder.  If desired, the degeneracy of the TE$_{111}$ modes can be tuned to a small fraction of a cavity bandwidth, although, as described in Ref.~\onlinecite{Chouinard.2025A}, it is better to deliberately separate the resonances by about one resonant bandwidth in order to resolve very weak TRSB signals.

The resonator is capped at the top and bottom by two circular end walls made of solid copper. In the center of each is a circular coupling iris, of typical diameter 4~mm.  Larger irises can be used, with the trade-off that increased coupling will lower cavity quality factor, or $Q$.
To maximize $Q$ in the low temperature experiment, the cavity side walls and end walls are either coated with electrolytically deposited Au or sputtered superconducting Nb. In the case of Nb, to prevent the superconducting coating from trapping flux following the application of a training field, thin lines have been scored across each superconducting surface to allow flux to easily enter and leave the interior of the resonator, although the lines of normal metal exposed in this process act to reduce cavity $Q$, which is then typically of the order of 50,000.  We are also careful to avoid creating rings of superconducting solder, so the resonator sidewall is glued into its brass housing using 2850~FT black epoxy for a permanent construction, or superglue for a version that can be disassembled.

\subsection{Interrogation with circularly polarized microwaves}

\begin{figure}[htb]
\includegraphics[width = 0.7 \columnwidth]{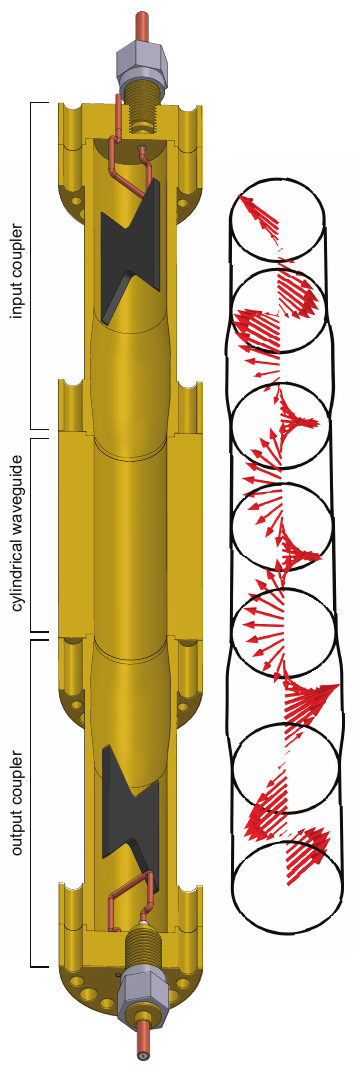}
\caption{Cutaway schematic (left) and finite element simulation (right) of a pair of quarter-wave tubes separated by a section of cylindrical waveguide, illustrating the conversion from linear to circular to 90-degree-rotated linear polarizations.  Arrows denote a snapshot of the local electric field polarization along the guide.}
\label{fig:finite_element}
\end{figure}

As outlined above, and discussed at length in Ref.~\onlinecite{Chouinard.2025A}, a powerful means of detecting and quantifying the degree of TRSB is to interrogate the resonator using circularly polarized microwaves. The scheme for doing so is shown in Fig.~\ref{fig:resonator_schematic}.  From left to right, microwaves propagate into the system, at a single frequency $\omega$, via the transverse electromagnetic (TEM) mode of a coaxial cable, and are converted into linear TE$_{11}$ polarizations of a cylindrical waveguide by a coax-to-waveguide adapter, shown in more detail in Fig.~\ref{fig:finite_element}.  The linearly polarized waveguide mode then enters an elliptical section of waveguide, with polarization oriented at 45$^\circ$ to the major and minor axes of the ellipse.  The elliptical section is precisely sized so that the component of the wave polarized along the slow axis of the ellipse emerges $\frac{\pi}{2}$ out of phase with the component polarized along the fast axis.  The elliptical section then acts as a quarter-wave tube (the microwave equivalent of an optical quarter-wave plate), converting the linearly polarized microwaves into circularly polarized microwaves, which for concreteness we assume to be $|+\rangle$ states.  The $|+\rangle$-polarized microwaves are then coupled into the resonator through the on-axis coupling iris on the left of the resonator, exciting a superposition of TE$_{111}$-derived resonator eigenmodes.  (Due to dissipation, these modes have finite bandwidth, so at minimum there is always some off-resonant excitation of each mode.)  Modulated by the amplitude and phase of the superimposed resonant modes, a small portion of the microwaves are then coupled out of the right-side coupling iris.  They are incident on a second quarter-wave tube, which converts the $|+\rangle$ component of the microwaves into a linear polarization that is able to pass through a second linear polarizer and coax-to-waveguide converter at the far right of the assembly.  Any $|-\rangle$ component that is coupled out of the resonator is rejected by the output coupler.

\begin{figure}[htb]
\includegraphics[width = 1.0 \columnwidth]{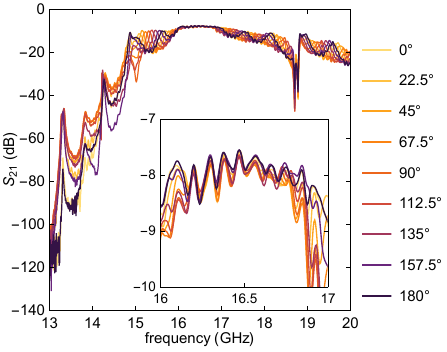}
\caption{Forward scattering parameter, $S_{21}(f)$, of the coupler test structure in Fig.~\ref{fig:finite_element}, demonstrating the weak dependence of transmission amplitude on relative coupler angle in the operating frequency range between 16 and 17~GHz.  The inset shows a close-up view of this frequency range.  The fine ripples, with approximately 100~MHz spacing, are not a property of the coupler test structure, but are due to standing waves in the much longer coaxial cables connecting the couplers to the microwave network analyzer.}
\label{fig:coupler_rotated}
\end{figure}

When the process is run in reverse, microwaves propagate from right to left.  In that case it is straightforward to show that the right-hand coupler now acts as a source of $|-\rangle$-polarized microwaves, allowing the resonator to be interrogated by circularly polarized microwaves of the opposite sense.\cite{Chouinard.2025A}  Any difference of the response of the resonator to $|\pm\rangle$ microwaves now shows up as a breaking of reciprocity --- i.e., a difference in the complex forward and reverse transmission amplitude through the resonator.  A complete derivation of this process is straightforward but somewhat involved, and is given in Ref.~\onlinecite{Chouinard.2025A}.

The process for launching and detecting circularly polarized microwaves is shown in more detail in Fig.~\ref{fig:finite_element}, in an experimental geometry used to test the quarter-wave character of the couplers, in which the central resonator is replaced by a section of cylindrical waveguide.  At the top end, we see that a microwave coaxial cable connects to a spark-plug-type SMA connector that screws into a mating housing in the end cap of the coupler.  The center conductor is soldered to a rectangular loop antenna, precisely sized to achieve good matching between the TEM mode of the 50~$\Omega$ coaxial cable and the TE$_{11}$ mode of the circular waveguide.  The loop antenna produces a linear polarization in which the rf $H$ field is perpendicular to the plane of the loop and the rf $E$ field is parallel.  In order to make sure this is the only polarization present in this part of the coupler --- i.e., to absorb unwanted reflections from the opposite end of the coupler, which would be rotated 90$^\circ$ by passing twice through the quarter-wave section --- the loop antenna is immediately followed by a bow-tie-shaped septum consisting of graphite-coated fiberglass.  The graphite coating, on the broad side faces of the bow tie, acts as a strong attenuator of electric fields polarized parallel to the plane of the bow tie, but allows the perpendicular polarization to pass with little attenuation.  The bow tie is oriented at 45$^\circ$ to the fast and slow axes of a section of elliptical waveguide, which converts the linearly polarized microwaves to circularly polarized microwaves as they enter the central, circular section of waveguide, which in our test setup is several guide wavelengths long.  This can clearly be seen in the finite element simulation shown alongside the schematic in Fig.~\ref{fig:finite_element}.  Microwaves exit the central section and pass into a second elliptical quarter-wave tube, which converts the microwaves back to linear polarization, albeit with plane of polarization rotated 90$^\circ$ from the input polarization. The linearly polarized microwaves pass through a second graphite-coated septum and back out into another coaxial cable via a second, identical loop antenna.  The combination of septum and loop antenna projectively selects the desired polarization.  To demonstrate that the circular polarizing couplers are indeed functioning as good quarter-wave tubes, we use this test setup to measure transmission response as a function of the relative orientation of the input and output couplers, as one is rotated with respect to the other.  If the couplers are producing purely circularly polarized microwaves, the transmission response will be independent of the relative orientation.  If, on the other hand, the microwaves have significant elliptical polarization, there will be strongly preferential transmission as a function of orientation angle.  As seen in Fig.~\ref{fig:coupler_rotated}, there is a broad band of frequencies between 16 and 17~GHz in which transmission changes by less that 1~dB on rotating relative coupler orientation from 0$^\circ$ to 180$^\circ$, confirming the proper functioning of the circular polarizers.

\begin{figure}[t]
\includegraphics[width = 0.75 \columnwidth]{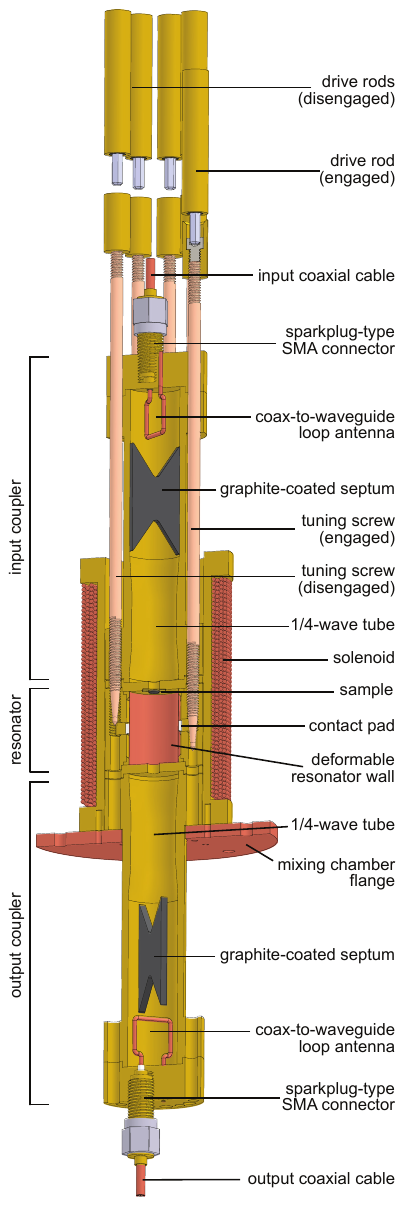}
\caption{Cutaway drawing of the TRSB resonator assembly with key components labeled, as described in the text.}
\label{fig:full_assembly}
\end{figure}

\begin{figure}[t]
\includegraphics[width = 0.6 \columnwidth]{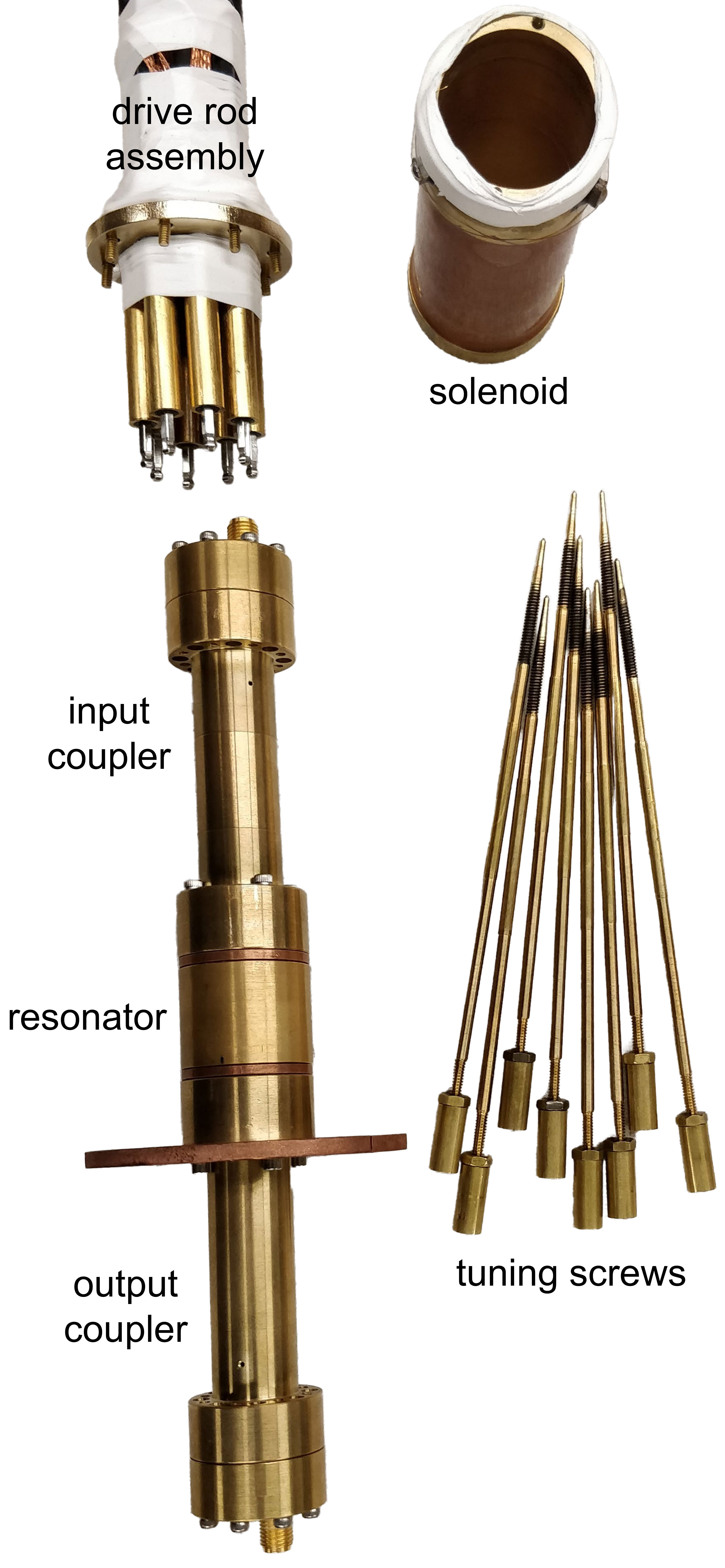}
\caption{Photograph of the drive-rod and resonator assemblies, showing tuning screws and solenoid removed.}
\label{fig:zoomed_assembly_photo}
\end{figure}

\subsection{Full experimental assembly}

The low temperature section of the experimental assembly is shown in cutaway in Fig.~\ref{fig:full_assembly}, which can be zoomed to reveal the finest details.  The couplers are now connected to either side of the central TE$_{111}$ resonator.  At the top end of the resonator is the sample, typically a mm-sized platelet single crystal mounted on a thin sapphire plate, positioned in the center of the upper end wall of the resonator at a local maximum of the transverse rf $H$ field.  Despite its position immediately in front of the coupling iris, the sample in no way ``blocks'' the entry of microwaves into the resonator, because it does not make electrical contact with the end wall, and presents a relatively minor perturbation to the resonator fields, which run transverse to the resonator axis at the sample position (i.e., the thin platelet sample is mounted in a low-demagnetizing-factor geometry with respect to the rf $H$ field).  

\begin{figure}[t]
\includegraphics[width = 0.53 \columnwidth]{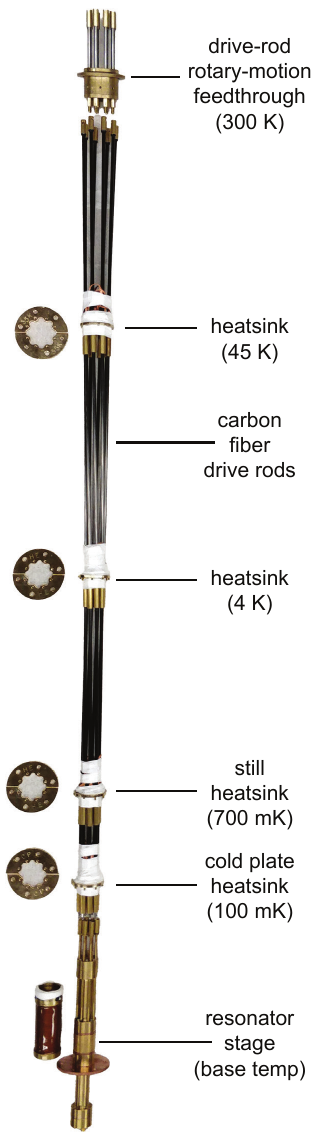}
\caption{Full experimental assembly showing the motion feedthrough that mounts into a KF40 port at the top flange of the LD400 BlueFors dilution refrigerator, which then connects to the eight pultruded carbon-fiber drive rods. The long drive rods are individually heat-sunk at the 40~K, 4~K, still and cold-plate flanges, then mate with the short drive rods that actuate the resonator deformation as shown in Fig.~\ref{fig:tunable_quadrupoles}. Each long drive rod is retracted once the desired deformation is set, eliminating any direct mechanical connection to the resonator while the experiment is running.}
\label{fig:full_insert_photo}
\end{figure}

The cross section through the resonator also reveals the small PTFE pads, which mediate between larger brass blocks and the BeCu resonator walls.  Two of the threaded tuning screws can be seen, one fully engaged, maximally deforming the wall at that location, and one fully disengaged.  Higher up, we see some of the pultruded carbon-fiber drive rods that actuate the eight tuning screws.  Each drive rod ends in a hexagonal key, which can be fully withdrawn from the corresponding keyway that it drives.  This allows two things.  First, the tuning screws can be rotated through an arbitrarily large range of angle --- even while we limit the drive rods, due to their thermal anchoring, to a safe operating range of about 70$^\circ$ of rotation --- by sequentially disengaging and re-engaging the drive rods while adjusting the tuning screws.  Then, when the quadrupolar distortions are properly adjusted, the drive rods can be completely retracted before starting the experiment, breaking any direct mechanical contact that could otherwise act as a source of intermittent perturbation to the resonator.  Although the drive rods are well anchored thermally, this also eliminates a potential pathway for heat conduction from higher temperatures, something that is always a concern when operating in the millikelvin regime.

The resonator is surrounded by a small superconducting solenoid, positioned so that it is centered on the sample, rather than on the resonator.  Our solenoid consists of 1495~turns of NbTi wire, and produces a training field up to 295~mT when energized by a current of 10~A.

The resonator assembly is bolted to a mounting flange that then attaches to the center of the mixing-chamber plate of a BlueFors LD400 dilution refrigerator.  The central location allows the drive rods to pass through a rotary-motion sliding-seal assembly mounted to a KF40 flange at the top of the dilution refrigerator vacuum can and down to the experiment, via an on-axis line-of-sight port.  

Figure~\ref{fig:zoomed_assembly_photo} shows a photograph of the low temperature components: the resonator and couplers; the tuning screws; the ends of the drive rods; and the superconducting solenoid.  Figure~\ref{fig:full_insert_photo} shows the full apparatus up to room temperature, including the rotary motion feedthrough going into the vacuum can, and the thermal anchoring of the drive rods, which is carried out using flexible copper braids, one attached to each drive rod at every source of intermediate-temperature cooling power in the dilution refrigerator (the 40~K and 4~K stages of the pulse-tube refrigerator, and the still and cold-plate flanges of the dilution unit).  The heat sinking is effective enough that a base temperature of 20~mK can readily be achieved at the resonator.

\section{Performance}

In this section we present several sets of test results, to both illustrate the operation of the TRSB resonator and gauge its ultimate sensitivity. In order to convert between the experimentally measured $\omega_\tau$  and more fundamental quantities such as  Kerr and Faraday angles, the geometry of the sample and cavity must be taken into account.  As shown in Ref.~\onlinecite{Chouinard.2025A}, this can be done using the theory of resonator perturbation, from which we obtain 
\begin{equation}
\theta_K = \frac{2 \omega_\tau}{\Gamma\,Z_0}\;,
\label{eq:Kerr_angle}
\end{equation}
where $\omega_\tau$ is an angular frequency (in rad/sec) and the resonator constant is 
\begin{equation}
    \Gamma = \frac{1}{2 \mu_0} \left.\int_S H^2 dS\middle/\right.\int_V H^2 dV\;.
    \label{eq:gamma1}
\end{equation}
The first integral is taken over the surface of the sample, the second over the volume of the cavity, and  $H(\mathbf{r})$ is the microwave magnetic field profile of the TE$_{111}$ mode.  We can always express these integrals in terms of a sample area, $A_\mathrm{sample}$, which for Kerr effect includes both the top and bottom faces of the sample, and an effective cavity volume
\begin{equation}
        V_\mathrm{eff} = \frac{1}{H_\mathrm{sample}^2} \int_V H^2 dV,
    \label{eq:effective_volume}
\end{equation}
allowing the resonator constant to be written as
\begin{equation}
    \Gamma = \frac{1}{2 \mu_0} \frac{A_\mathrm{sample}}{V_\mathrm{eff}}\;. 
    \label{eq:gamma2}
\end{equation}
\begin{figure}[t]
\includegraphics[width = 1.0 \columnwidth]{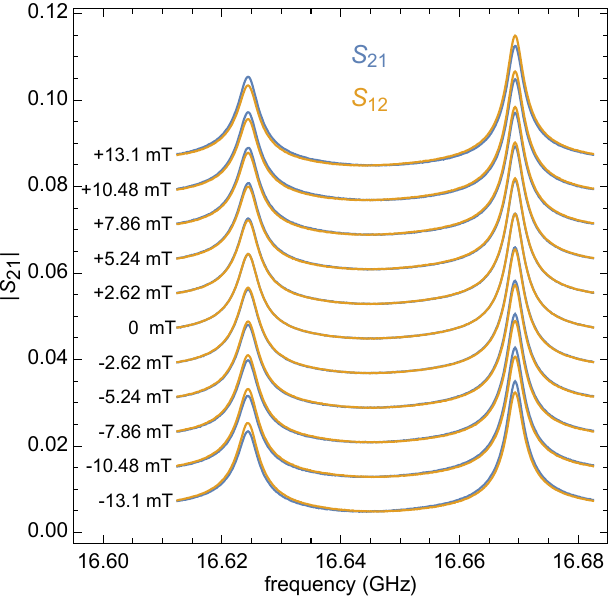}
\caption{Forward ($S_{21}$) and reverse ($S_{12}$) resonance traces taken on a 10~mm$^2 \times 0.65$~mm ferrite test sample as a function of magnetic field, showing field-induced breaking of reciprocity.  (Traces have been offset for clarity.)}
\label{fig:YIG_freq_sweeps}
\end{figure}
In our system, the sample is typically a thin platelet, small compared to the cavity cross section, located parallel to and at the center of one of the end walls, where the transverse magnetic field is maximum.  In that case, for the TE$_{111}$ mode of a cylindrical cavity,\cite{Chouinard.2025A}
\begin{equation}
        V_\mathrm{eff} = 0.239\,V_\mathrm{cavity} \left(1 + 0.343\frac{d^2}{a^2}\right)\;.
        \label{eq:effective_volume_TE111}
\end{equation}
In our implementation, the cavity has height $d = 18$~mm and radius $a = 6$~mm, leading to an aspect ratio $d/a = 3$ and an effective volume $V_\mathrm{eff} = 0.978~V_\mathrm{cavity}$.

Substituting into Eq.~\ref{eq:Kerr_angle}, we obtain
\begin{equation}
\theta_K = 4 \omega_\tau\frac{\mu_0}{Z_0} \frac{V_\mathrm{eff}}{A_\mathrm{sample}}= \frac{4 V_\mathrm{eff}}{c~A_\mathrm{sample}} \times \omega_\tau\;,
\end{equation}
where $c$ is the speed of light in vacuum.

\begin{figure}[t]
\includegraphics[width = 1.0 \columnwidth]{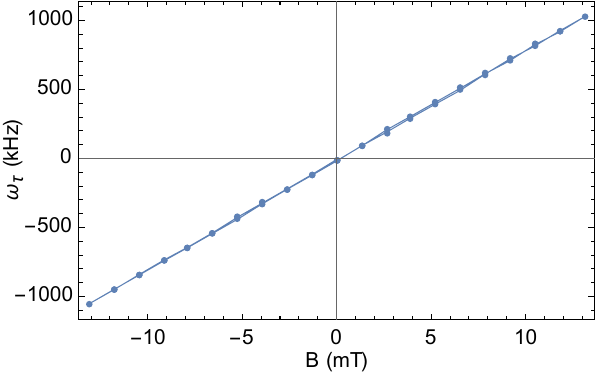}
\caption{A hysteresis loop showing field-induced breaking of time-reversal symmetry in a 0.65-mm-thick ferrite slab, obtained from fits to the 16.6~GHz frequency sweeps in Fig.~\ref{fig:YIG_freq_sweeps}.  Experimentally measured $\omega_\tau$ is shown on the left axis (as a periodic frequency, in MHz) and converted to Faraday angle on the right axis.}
\label{fig:YIG_field_sweeps}
\end{figure}

In the first instance, we show data in which TRSB is visible to the naked eye: room-temperature measurements of a 10~mm$^2 \times 0.65$-mm-thick slab of ferrite taken from a microwave isolator. Ferrites are soft ferromagnetic insulators in which breaking of time-reversal symmetry is readily induced by the application of a weak magnetic field, making them ideal for nonreciprocal devices such as isolators and circulators. Electromagnetic waves propagate through the ferrite with little attenuation, meaning that instead of a Kerr effect, which results from the nonreciprocal phase shift when LCP and RCP waves are \emph{reflected} from a surface, the ferrite exhibits a Faraday effect, in which the Faraday rotation angle is $\theta_F = \phi_\mathrm{nr}/2$, where $\phi_\mathrm{nr}$ is the nonreciprocal phase shift between LCP and RCP waves \emph{transmitted} through the sample.  (For a single pass, $\theta_F$ therefore takes the same form as $\theta_K$, allowing Eq.~\ref{eq:Kerr_angle} to also be used to calibrate a Faraday rotation measurement.)   During each cycle of the cavity, microwaves transmitted through the ferrite are subsequently reflected from the cavity end wall and pass through the ferrite sample a second time, doubling the nonreciprocal phase shift that is acquired (and doubling the effective path length).  The observation of a nonzero net phase shift during the back-and-forth transit is an important signature of nonreciprocity, confirming a true Faraday effect due to circular birefringence.  By contrast, phase shifts arising from linear birefringence are undone when reflected back through the material, maintaining reciprocity.  

Forward ($S_{21}$) and reverse ($S_{12}$) frequency sweeps are shown in Fig.~\ref{fig:YIG_freq_sweeps} for the ferrite sample, at a set of fixed fields ranging from -13~mT to +13~mT. In zero field, the forward and reverse traces are the same: i.e., the resonator response is reciprocal.  As field is applied, the resonance traces become progressively less reciprocal, and the sign of the effect changes when the field direction is reversed.  By fitting to the resonance traces and applying Eq.~\ref{eq:ratiometric_measure}, $\omega_\tau$ is obtained for each pair of forward and reverse sweeps. $\omega_\tau$ is then plotted vs.\ field in Fig.~\ref{fig:YIG_field_sweeps} and converted to Faraday angle, yielding a Faraday rotation of 1.3~rad/tesla for the 0.65~mm thick sample, which is typical for a ferrite in the microwave range.\cite{Hogan.1952}

\begin{figure}[t]
\includegraphics[width = 1.0 \columnwidth]{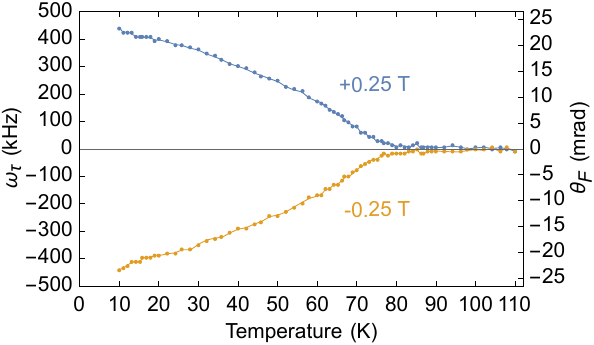}
\caption{Temperature sweeps in the insulating van der Waals ferromagnet CrGeTe$_3$, in magnetic fields $B = \pm 0.25$~T.  Experimentally measured $\omega_\tau$ is shown on the left axis (as a periodic frequency, in MHz) and converted to Faraday angle on the right axis.}
\label{fig:CeGeTe3}
\end{figure}

We next show temperature-sweep data taken on a small single crystal of the van der Waals ferromagnet CrGeTe$_3$, acquired using a version of the TRSB resonator mounted in a helium flow cryostat.  CrGeTe$_3$ has a Curie temperature $T_c \approx 66$~K\cite{Lee2024} and, since it is a magnetic insulator, the appropriate measure of TRSB is again the Faraday angle $\theta_F$.  Figure~\ref{fig:CeGeTe3} shows $\theta_F$ measured as a function of temperature, in magnetic fields of $B = \pm 0.25$~T.  The applied field induces a nonzero moment above $T_c$, with TRSB visible at temperatures below about 80~K.

\begin{figure}[t]
\includegraphics[width = 1.0 \columnwidth]{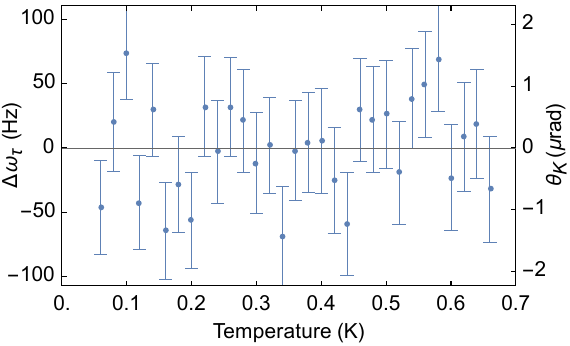}
\caption{Low temperature performance, showing a null temperature sweep taken overnight in the dilution refrigerator.  The temperature trace is flat, indicating an absence of systematic error.  $\omega_\tau$ is shown on the left axis (as a periodic frequency in Hz) and converted to Kerr angle on the right axis assuming a sample size of 2~mm$\times$2~mm.  The measurement uncertainty, $\delta \omega_\tau = \pm 39$~Hz, corresponds to a Kerr-angle resolution of $\delta \theta_K = 810$~nanoradian.}
\label{fig:low_temperature_trace}
\end{figure}

As an illustration of the low temperature stability and sensitivity, in Fig.~\ref{fig:low_temperature_trace} we show data acquired in an optimized, Nb-coated resonator, mounted in the BlueFors LD400 dilution refrigerator.  In this temperature range, it is critically important that we avoid self-heating from the microwaves, so we use an output power of -55~dBm from the microwave network analyzer.  This undergoes an additional attenuation of 15~dB by the combination of cryogenic microwave cables and cavity coupling iris, so that the input power to the microwave resonator is -70~dBm = 100~pW.  Unlike other low temperature microwave spectroscopy experiments,\cite{Truncik:2013hr} it is not possible  to use low noise preamplifiers in the TRSB system to boost signal level and lower system noise floor: the amplifiers cannot be inserted into the main measurement loop, as we have to be able to measure in both the forward and reverse direction  to obtain $S_{21}$ and $S_{12}$; similarly, experience has shown that if we attempt to add amplification at the microwave network analyzer input ports, between the signal separation and receiver stages, then calibration is compromised to the point where it is not possible to meaningfully detect changes in reciprocity.  So data are typically taken without any amplification, making the experiments very challenging.  Nevertheless, we are able to achieve a measurement uncertainty of $\delta \omega_\tau = \pm 39$~Hz in a temperature sweep taken overnight in the dilution refrigerator, corresponding to a Kerr-angle resolution $\delta \theta_K = 810$~nanoradian for a  2~mm$\times$2~mm sample.  This resolution could be improved by averaging the signal for a longer period, or by measuring a finely divided sample with larger surface area.  (To give an indication of the utility of the latter effect, a 1~mm$^3$ crystal powdered into 100~$\mu$m grains would lead to a factor of 10 improvement in resolution; 10~$\mu$m grains would lead to a factor of 100 improvement.) Importantly, we see in Fig.~\ref{fig:low_temperature_trace} that the temperature trace is completely flat within experimental uncertainty, allowing the expected onset of spontaneous TRSB at the superconducting transition to be resolved unambiguously.

\begin{acknowledgments}
We acknowledge A.~Kapitulnik for suggesting that TRSB be measured with microwaves, and acknowledge initial attempts by K.~J.~Morse to realize that goal.  We specifically thank E.~Mun for providing the CrGeTe$_3$ test sample.  We are grateful to J.~S.~Dodge, E.~Girt, P.~J.~Hirschfeld, V.~Mishra, E.~Mun and J.~E.~Sonier for useful discussions and technical assistance.  We thank the SFU Silicon Quantum Technology Lab for extensive access to their BlueFors dilution refrigerator and microwave electronics. Financial support for this work was provided by the Natural Science and Engineering Research Council of Canada. 
\end{acknowledgments}

\section*{Data Availability Statement}

The data that support the findings of this study are available from the corresponding author upon reasonable request.



%

\end{document}